\documentclass[preprint]{aastex}
\usepackage{url}

\RequirePackage{color}
\newcommand{\emaila}{m.bocchi@imperial.ac.uk}

\begin{document}

\title{Numerical study of jets produced by conical wire arrays on the Magpie pulsed power generator}
\shorttitle{Simulations of jets from conical arrays}
\shortauthors{Bocchi et al.}


\author{M.~Bocchi} 
\email{\emaila}
\and 
\author{J.~P.~Chittenden}
\affil{The Blackett Laboratory, Imperial College London, London, UK}
\and
\author{A.~Ciardi}
\affil{LERMA, Universit\'e Pierre et Marie Curie, Observatoire de Paris, Meudon, France}
\affil{\'Ecole Normale Sup\'erieure, Paris, France. UMR 8112 CNRS}
\and
\author{F.~Suzuki-Vidal}
\and
\author{G.~N.~Hall}
\and
\author{P.~ de Grouchy}
\and
\author{S.~V.~Lebedev}
\affil{The Blackett Laboratory, Imperial College London, London, UK}
\and
\author{S.~C.~Bott}
\affil{Center for Energy Research, University of California, San Diego, CA, USA}
\email{The final pubblication will be available on Springer.}


\begin{abstract}
The aim of this work is to model the jets produced by conical wire arrays on the MAGPIE generator, and to 
design and test new setups to strengthen the link between laboratory and astrophysical jets.
We performed the modelling with direct three-dimensional magneto-hydro-dynamic numerical simulations using 
the code GORGON. We applied our code to the typical MAGPIE setup and we successfully reproduced the experiments. 
We found that a minimum resolution of $\sim 100$ $\mu m$ is required to retrieve the unstable character of the jet. 
We investigated the effect of changing the number of wires and found that arrays with less wires produce more 
unstable jets, and that this effect has magnetic origin. Finally, we studied the behaviour of the conical array 
together with a conical shield on top of it to reduce the presence of unwanted low density plasma flows. 
The resulting jet is shorter and less dense. 
\end{abstract}



\section{Introduction}


Although important improvements have been achieved in the comprehension of astrophysical jets, several questions 
remain open, including jet formation, propagation in an external medium and survival to potentially disruptive 
instabilities \citep{bellan09,hardee04}. Jets produced in the laboratory are scalable to astrophysical conditions, as they are 
characterised by dimensionless parameters in a similar range to Young Stellar Object (YSO) jets. In particular, 
jets produced by conical wire arrays are especially suitable to study the interaction of the jet with an ambient 
medium \citep{lebedev05,ciardi08}.
In this framework we produced laboratory jets on the MAGPIE pulsed power generator at Imperial College, 
London \citep{mitchell96} using different setups: conical and radial wire arrays \citep{lebedev05} and radial foils \citep{ciardi09}. The recent upgrade 
of MAGPIE and other pulsed power facilities (notably the "Z" machine in Sandia, USA) will allow the study of new 
physical regimes for jets. The work presented here aims to carefully model  the experiments of jets from conical 
wire arrays on MAGPIE in order to understand the physics and help the design of new experimental setups and 
diagnostics. Direct numerical simulations are employed for the modelling, using the three-dimensional (3D) resistive 
magneto-hydro-dynamic (MHD) code GORGON \citep{chittenden04,ciardi07}. 

\subsection{Conical wire arrays}
\begin{figure}[t]
\begin{tabular}{c@{\hspace{0.02\textwidth}}c}
\includegraphics[width=0.21\textwidth]{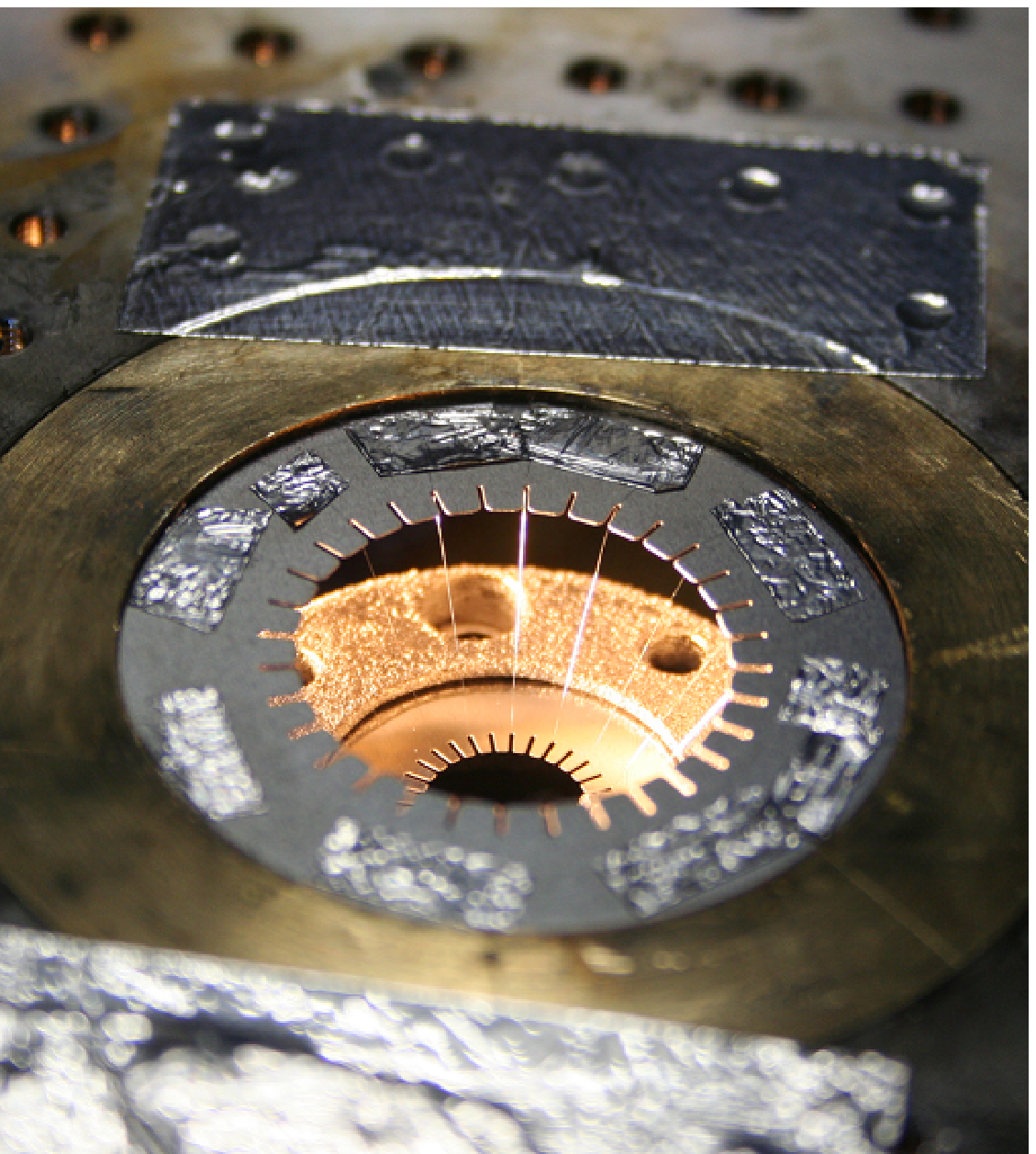} &
\includegraphics[width=0.21\textwidth]{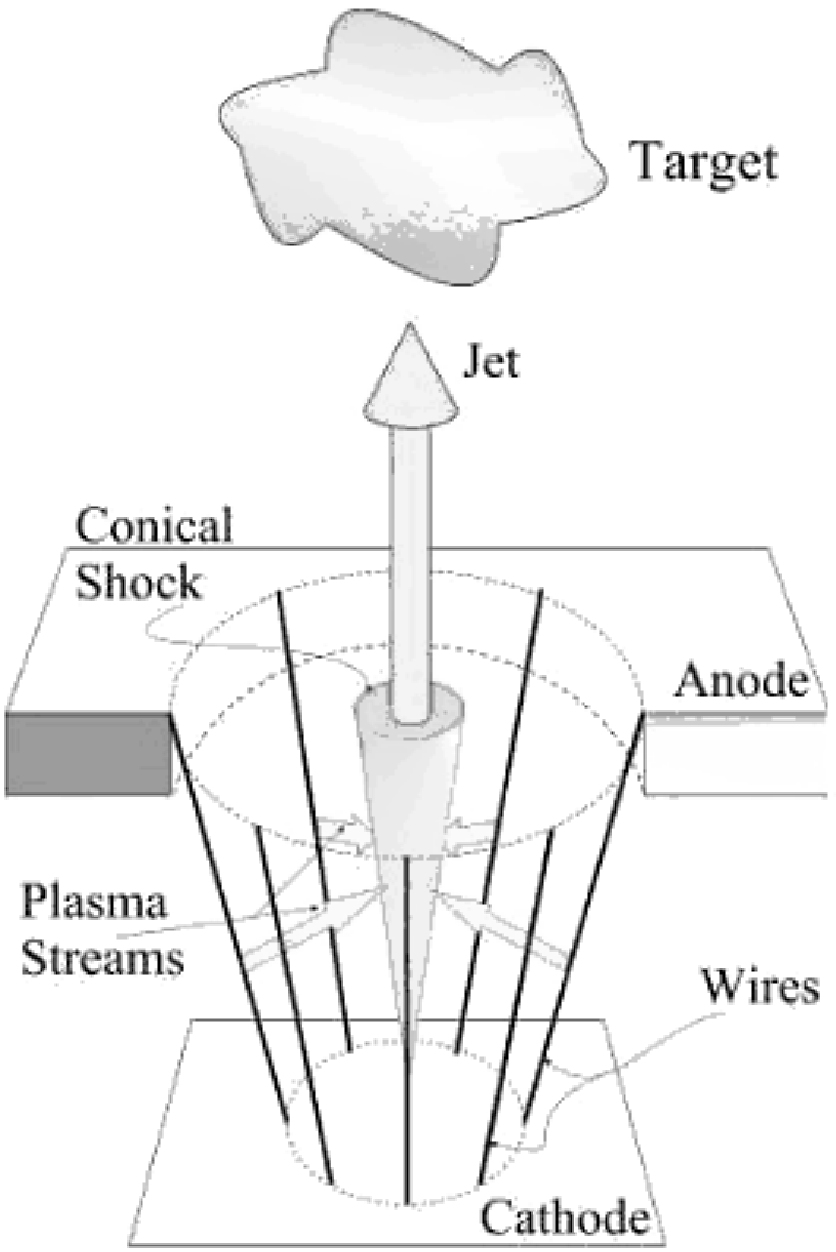} \\
\end{tabular}
\caption{%
\label{fig:cartoon}
{\it Left Panel:} Picture of the experimental setup. {\it Right Panel:} Scheme of a typical conical array setup
} 
\end{figure}

A conical wire array is composed by a set of wires placed in a conical shape (see Fig.~\ref{fig:cartoon}) and 
connected to a pulsed power generator. As the electric current passes through the array, the wires are 
heated and transformed into plasma. The current also produces a global toroidal magnetic field which accelerates 
the ablated plasma in a direction perpendicular to the wires. The plasma streams are directed towards the array 
axis where a conical shock is formed. 
While the component of the velocity perpendicular to the shock surface is reduced by the large shock compression 
factor ($\sim 20$ - $40$),  the parallel component of the velocity remains continuous across the shock. In addition, 
since the conical shock has a small opening angle (with respect to the conical array angle), the post-shock 
flow velocity is approximately vertical, and results in a collimated jet that propagates axially.

\subsection{Plan}
In the next section (Sec.~\ref{sec:model}) we will briefly introduce the model and the numerical setup. We will present 
the results of the numerical simulations in section \ref{sec:results}. To start, we will compare the simulations of a 
standard case with new data from MAGPIE experiments. In addition, we will present our results on a possible way 
to improve jet experiments using a conical shield above the conical array. In the last section (Sec.~\ref{sec:conclusions}) we will 
draw the conclusions.

\section{Model and numerical setup}
\label{sec:model}

\subsection{Numerical code}
To carry out our simulations we used the numerical code GORGON \citep{chittenden04,ciardi07}, a Van Leer type 3D numerical 
code designed to integrate the MHD system of equations in a single fluid approximation. The 
energy equations of ions and electrons are solved separately. Ohmic heating and optically thin radiation losses 
are also considered. Electromagnetic fields are followed by diffusing and advecting the vector potential 
\textbf{A}. Zones 
under a cut-off density ($\rho_{vac}=10^{-4}$ $Kg/m^{3}$) are considered a computational ''vacuum'', where only the 
wave equation 
for \textbf{A} is solved.

\subsection{Initial setup and numerical methods}

The numerical setup has been chosen in order to mimic the recent experiments carried out on the MAGPIE facility in 
London. The numerical simulations employed a simplified initial setup consisting of two electrode plates 
connected by a variable number of equally spaced wires, as seen in Fig.~\ref{fig:cartoon}.
In the reference case studied in this paper, the cathode diameter is $d_c=12$ $mm$, the anode diameter
is $d_a=21$ $mm$, and the array length 
(the distance between cathode and anode) is $l=12$ $mm$. These values correspond to an inclination angle of 
$30.26$ degrees. Either $16$ or $32$ wires were simulated.
We modelled the electric current provided by the MAGPIE generator with the following expression:
\begin{equation}
\label{eq:current}
I(t)=I_0\cdot sin^2(\pi t/2\tau),
\end{equation}
where $I_0=1.4\cdot 10^6$ is the peak current in Ampere and $\tau=250\cdot10^{-9}$ is the peak time in seconds.
Computationally, this current is translated into appropriate conditions for the magnetic field at the boundaries 
of the computational domain.

\section{Results}
\label{sec:results}

\begin{figure*}[t]
\begin{tabular}{c@{\hspace{0.02\textwidth}}c@{\hspace{0.02\textwidth}}c}
\includegraphics[width=0.31\textwidth]{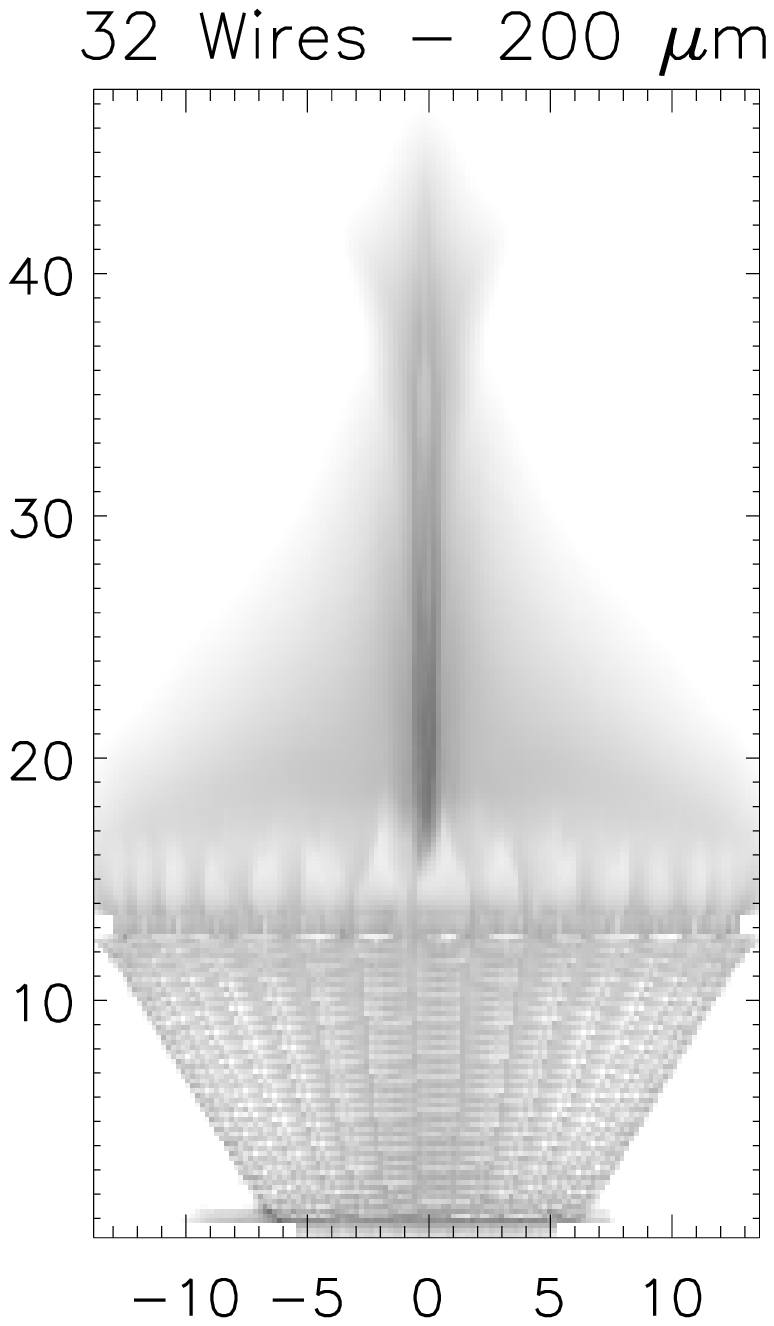} &
\includegraphics[width=0.31\textwidth]{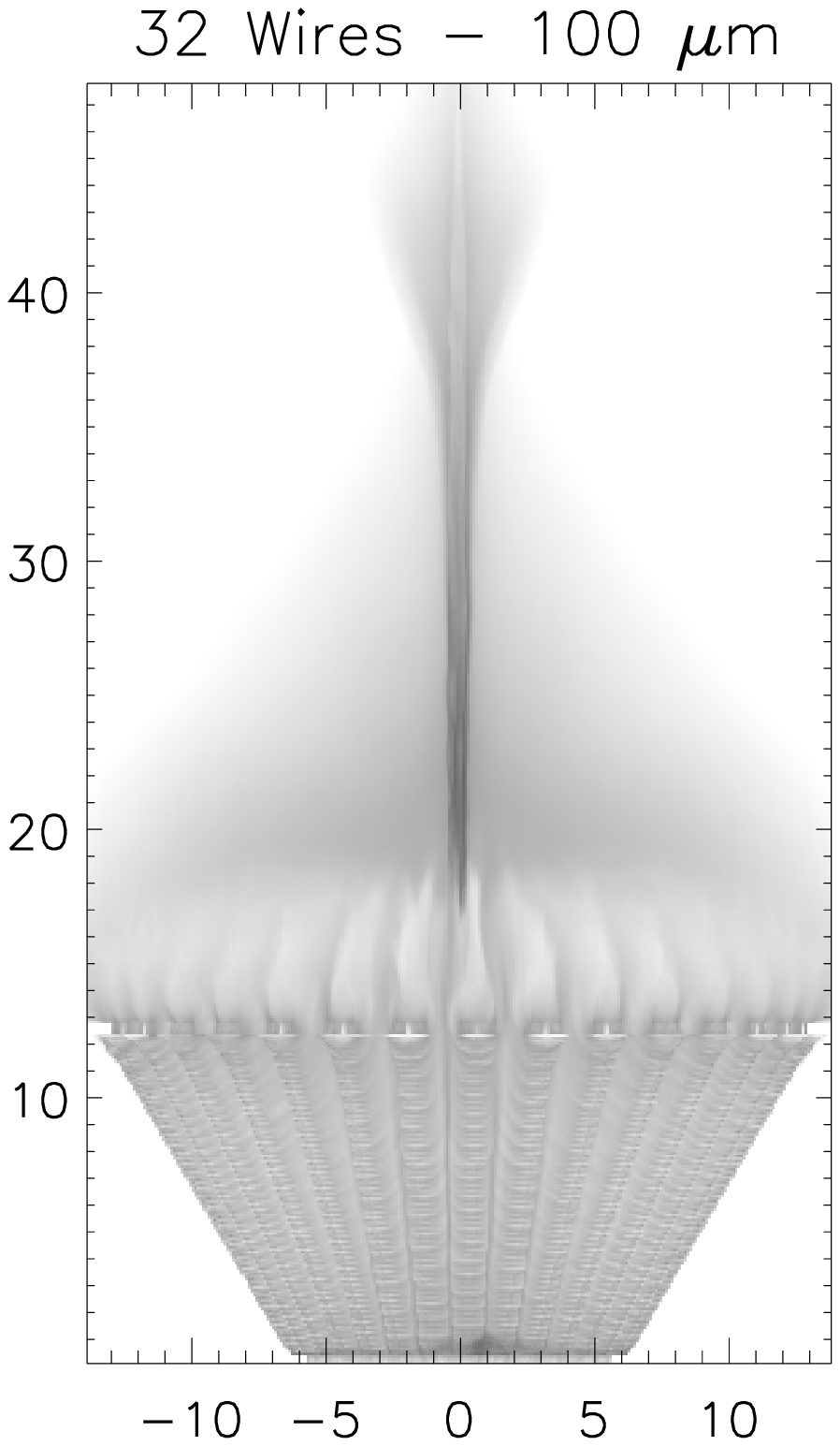} &
\includegraphics[width=0.31\textwidth]{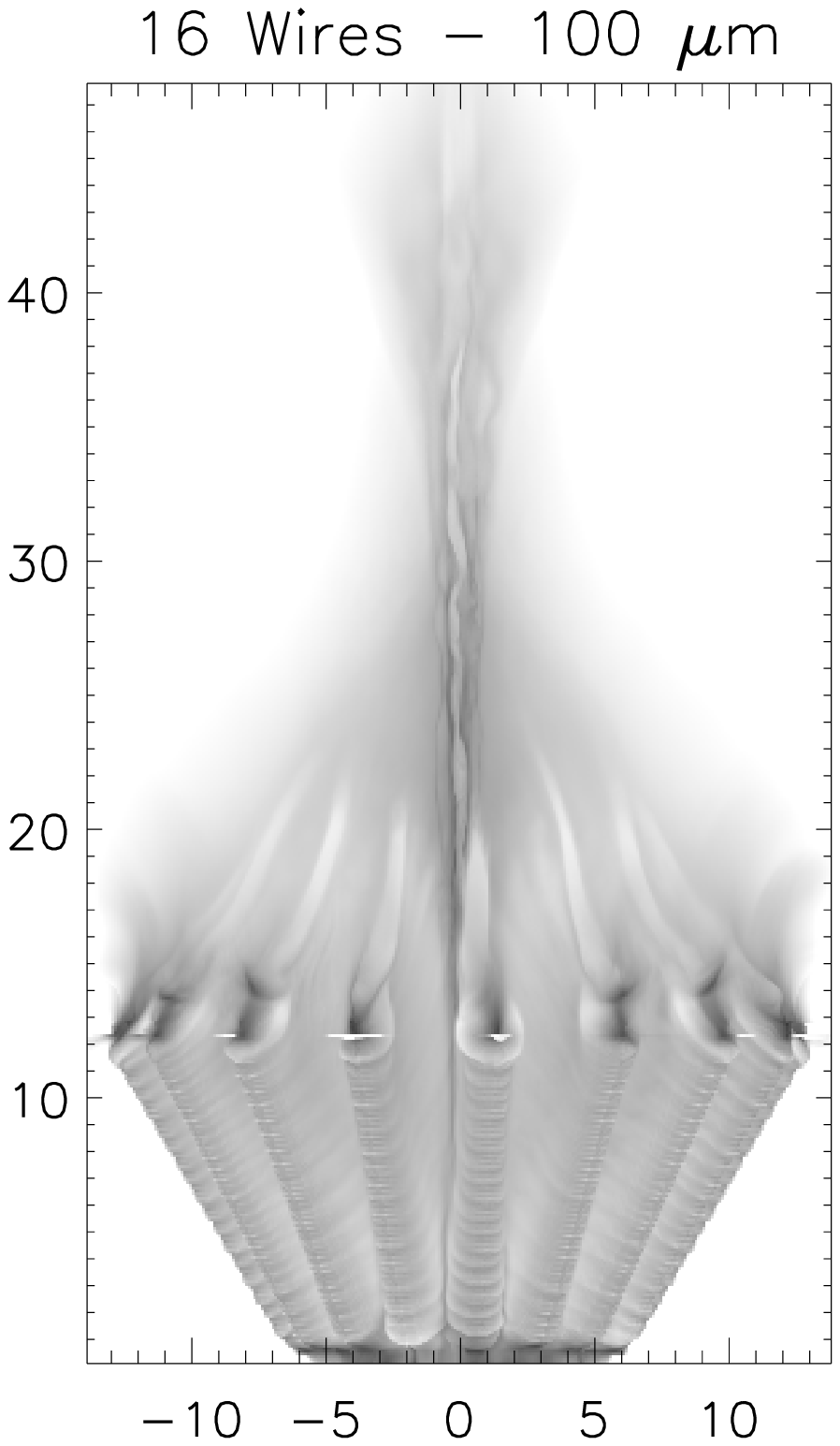} \\
\end{tabular}
\caption{%
\label{fig:emission}
Emission maps of three different simulations at $350$ ns. Darker shades correspond to higher emission. 
The axis are in $mm$ units. Only part of the computational box is shown in order to highlight the jet and 
structures 
{\it Left Panel:} $32$ wires, 
$200$ $\mu m$ resolution. {\it Central Panel:} $32$ wires, $100$ $\mu m$ resolution. 
{\it Right Panel:} $16$ wires, $100$ $\mu m$ resolution} 
\end{figure*}

\begin{figure*}[t]
\begin{center}
\includegraphics[width=0.30\textwidth]{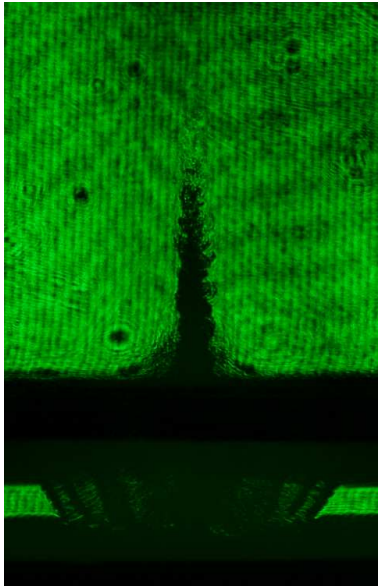}
\end{center}
\caption{%
\label{fig:shadowgram}
Laser shadowgram of the shot $s0301\_10$ taken at $355$ $ns$. Array of $16$ tungsten wires of $18$ $\mu m$ 
diameter 
} 
\end{figure*}

\begin{figure}[t]
\begin{tabular}{c@{\hspace{0.02\textwidth}}c}
\includegraphics[width=0.45\textwidth]{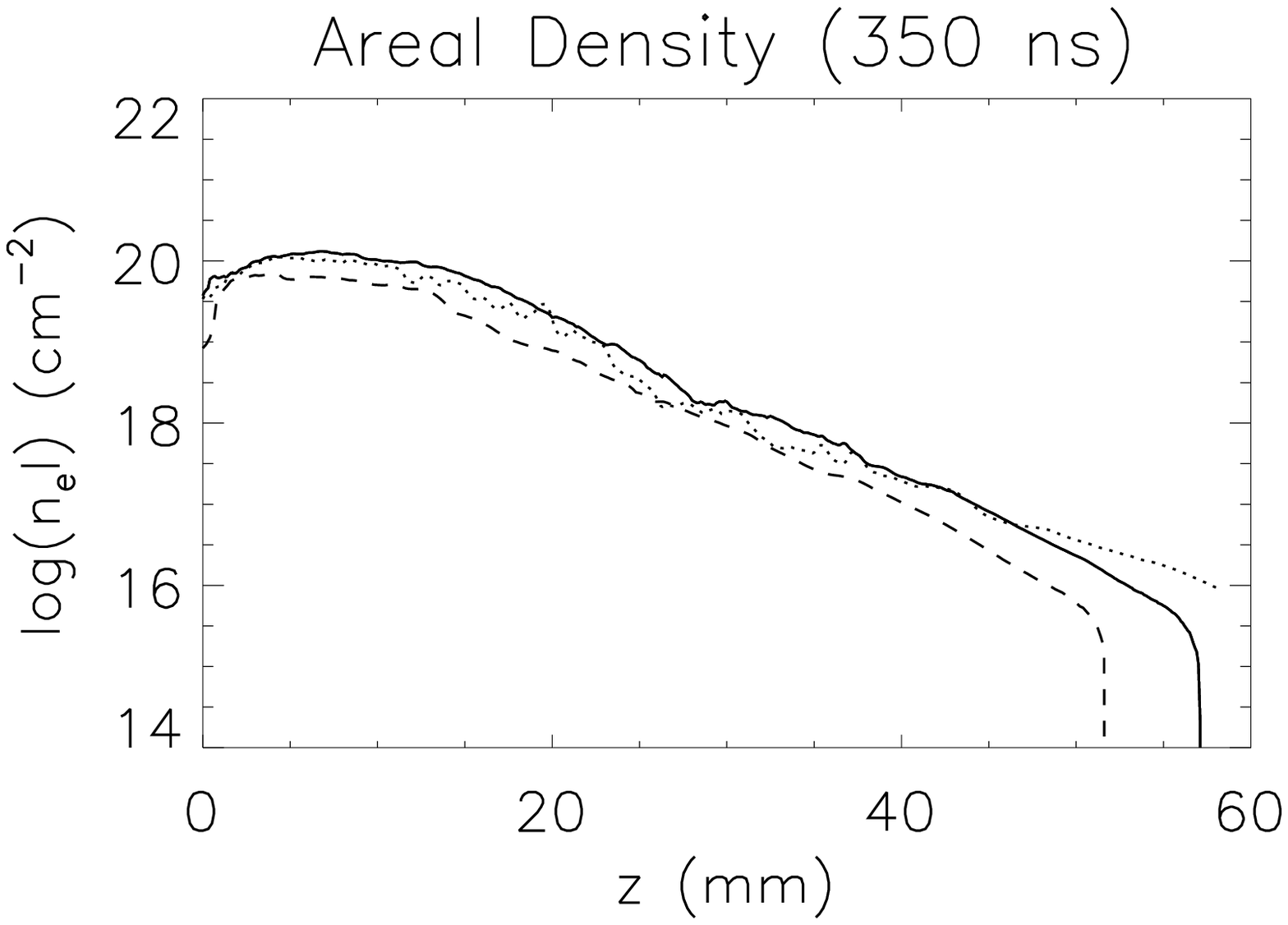} &
\includegraphics[width=0.45\textwidth]{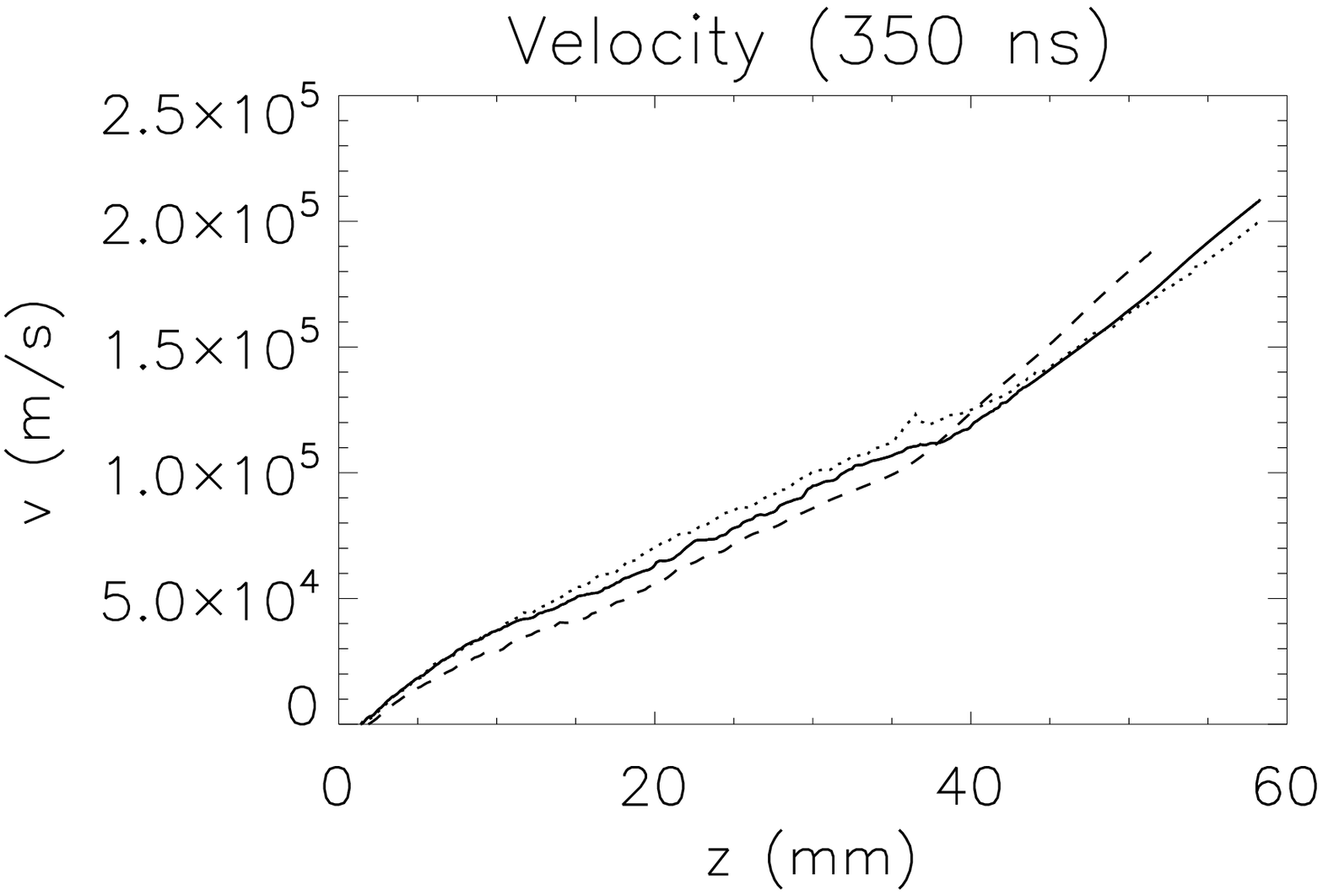} \\
\end{tabular}
\caption{%
\label{fig:cuts_compare}
Cuts along the array axis of the areal electron number density ($n_el$) ({\it Left Panel}), 
and of the axial component of the velocity ({\it Right Panel}). Data taken from various simulations 
at $350$ ns. {\it Solid Line:} $32$ wires, $100$ $\mu m $ resolution. 
{\it Dashed Line:} $32$ wires, $200$ $\mu m$ resolution. {\it Dotted Line:} $16$ wires, $100$ $\mu m$ resolution
} 
\end{figure}

\begin{figure}[t]
\begin{tabular}{c@{\hspace{0.02\textwidth}}c}
\includegraphics[width=0.45\textwidth]{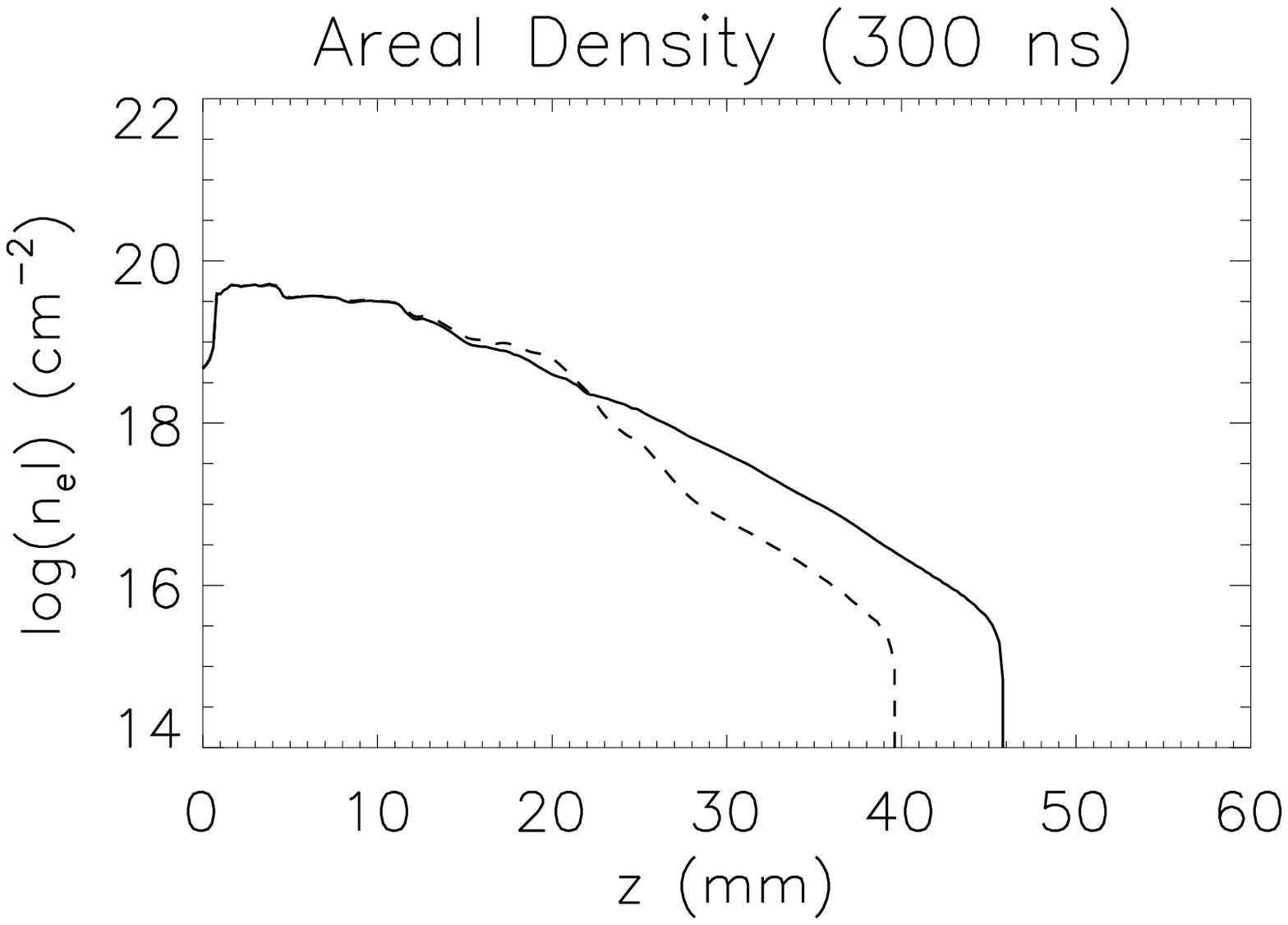} &
\includegraphics[width=0.45\textwidth]{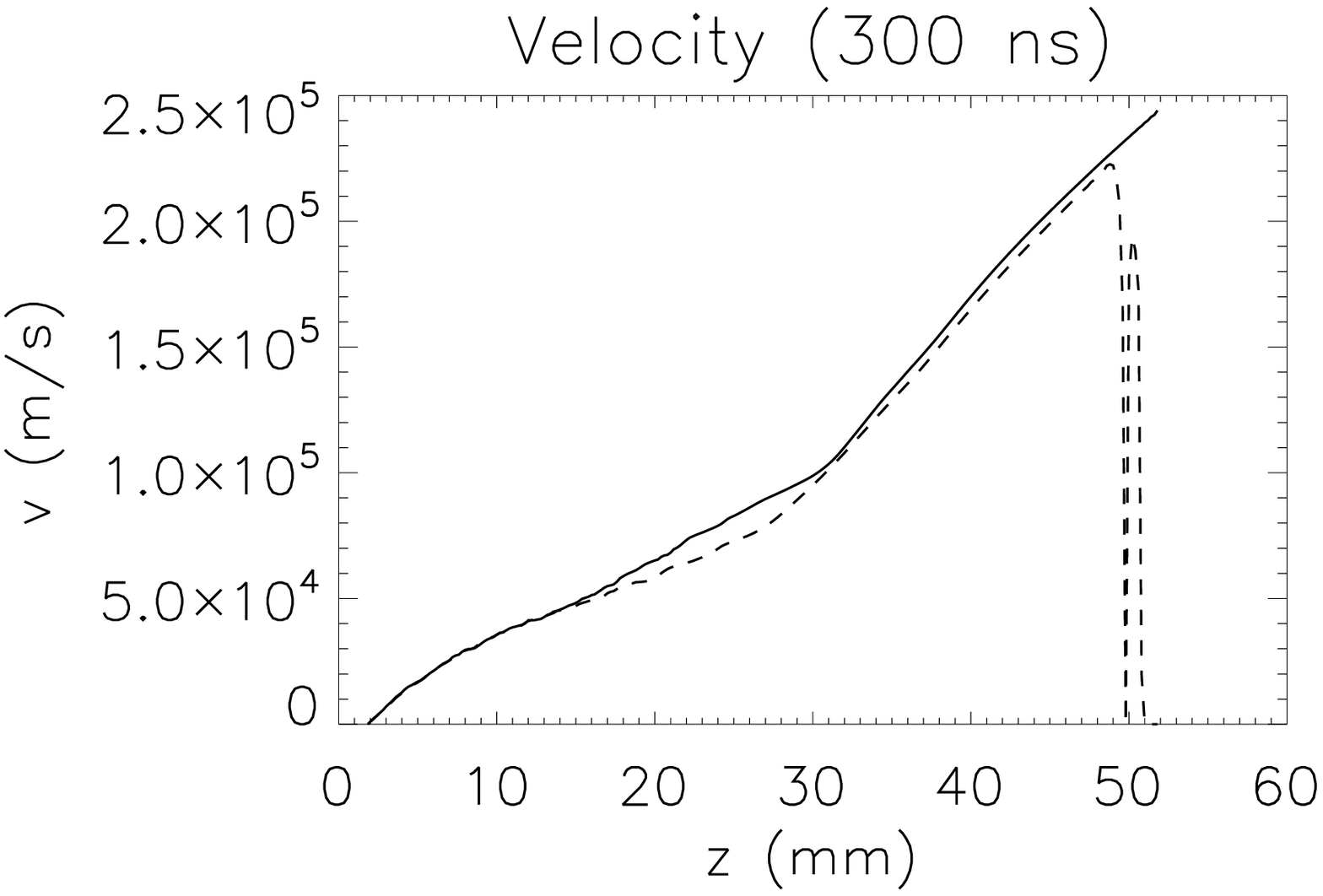} \\
\end{tabular}
\caption{%
\label{fig:cuts_cone}
Cuts along the array axis of the areal electron number density ($n_el$) ({\it Left Panel}), 
and of the axial component of the velocity ({\it Right Panel}). Data taken from the $32$ wires 
simulation at $200$ $\mu m$ resolution, at $300$ ns. {\it Solid Line:} No conical shield. 
{\it Dashed Line:} Conical shield with $8$ $mm$ aperture diameter} 
\end{figure}

\begin{figure}[t]
\begin{center}
\includegraphics[width=0.33\textwidth]{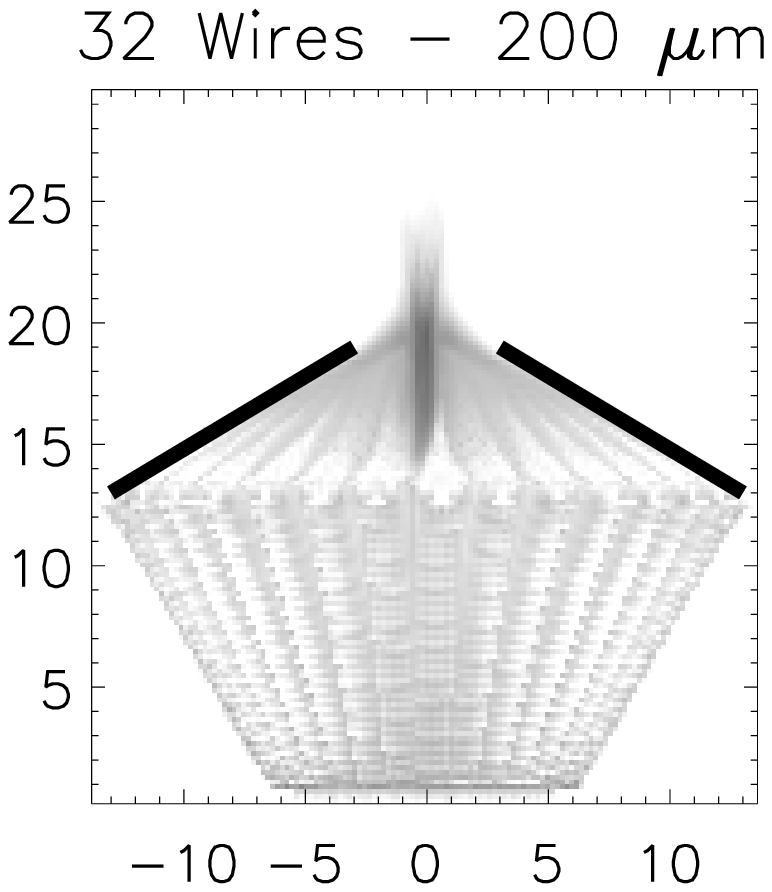}
\end{center}
\caption{%
\label{fig:cone_emission}
Emission map of the array simulation with conical shield at $300$ ns. Darker shades correspond to higher emission. 
The axis are in $mm$ units. Only part of the computational box is shown in order to highlight the jet and 
structures. The thick black lines represent a section of the conical shield 
} 
\end{figure}

\subsection{Reference Case}
\label{sec:ref_case}

We investigated numerically the behaviour of a conical wire array of $16$ and $32$ wires 
inclined by $30.26$ degrees with respect to the axis. The resolution of the simulations was varied between 
$100$ and $200$ $\mu m$.
The self emission maps in Fig.~\ref{fig:emission} show evident differences between 
the various cases.

First, the jet in the high resolution simulations is more turbulent and seems unstable.
Indeed, while the jet in the low resolution simulations is perfectly smooth, the high resolution 
counterpart presents non axisymmetric features. This is particularly evident for the $16$ 
wires case. The unstable character of the jet is observed in the experiments, as seen in 
the laser shadowgram in Fig~\ref{fig:shadowgram}. 
In Fig.~\ref{fig:cuts_compare} we plot the electron density per unit area ($n_e l$) and plasma velocity along 
the array axis, and draw a comparison of the various cases.
The $n_e l$ is systematically lower and smoother in the low resolution 
simulation, but the shape of the curve is the same. Velocity and temperature (not shown) are similar in 
both high and low resolution simulations. In contrast, the current density along the jet axis is compatible 
only in the lower part of the jet up to $z\sim 20$ $mm$, but is sensibly 
higher in both $100$ $\mu m$ resolution simulations above this height.

The jet diameter has been measured in the experiments to be $1$ - $2$ $mm$. 
At the same time, numerical codes typically require $3$ - $5$ grid cells to resolve steep gradients, 
e.g. as the radial density gradients present at the surface of the jet. Therefore, a resolution 
of $\sim 100$ $\mu m$, or finer, is needed to properly reproduce the jet profile.
However, the general trend of the physical variables along the jet axis is roughly preserved in the 
low resolution runs.  

Second, the jet produced by  the $16$ wires array appears to be much more unstable than the one in the $32$ 
wires array. In addition, in the $16$ wires simulation, the plasma streams at the height of the anode plate 
are bended inwards. 
The mass distribution around the jet is clearly less smooth for lower number of wires, resulting in 
larger perturbations on the beam.
Moreover, a careful examination of the magnetic field structure reveals that, in the case of $16$ 
wires, the magnetic field can easily penetrate the inter-wire gap and dominate the dynamics in the volume 
around the jet. Moreover, a small but not negligible vertical component of the magnetic field is observed. 
This picture is confirmed by the value of the plasma $\beta$ in the region immediately around the jet, 
which is less than unity up to 
$z\sim 27$ $mm$. Conversely, in the $32$ wires simulation, the plasma $\beta$ is  smaller than unity only up to 
$z\sim 15$ $mm$, just few millimetres above the anode. This suggests that the difference observed in the emission 
maps is partially due to magnetic effects.

\subsection{Conical Shield}
\label{sec:cone}

Conical wire arrays can be used to study the interaction of the jet with a target. 
Properly chosen targets allow for experiments relevant to the astrophysical scenario of a YSO jet interacting 
with the interstellar medium.  
However, as can be argued from Fig.~\ref{fig:emission}, the laboratory jet is surrounded by low density 
material, which can interact with the target 
and change the properties of it before the actual interaction.
In order to reduce the importance of such low density plasma, we tested a conical shield positioned above the 
array. The shield is designed to be perpendicular to the wires to prevent material to flow vertically above it. 
The thick black lines in Fig.~\ref{fig:cone_emission} illustrate the position of the conical shield with respect 
to the wire array.

The results presented in this section refer to a $32$ wires array simulation with $200$ $\mu m$ 
resolution. As explained in the previous section (Sec.~\ref{sec:ref_case}), such a resolution is not sufficient 
to reproduce all the details of the experiment, but gives insights on the behaviour of the physical quantities, in 
particular of density and velocity. 
In Fig.~\ref{fig:cuts_cone} we plot the velocity and $n_e l$ along the jet axis at $300$ $ns$ for the same array 
with and without the conical shield. It is evident 
that the presence of the shield does not affect the velocity of the flow by a considerable amount. However, the 
$n_e l$ plot reveals several differences between the two cases. In the presence of a conical shield, the jet is 
shorter and notably less dense. The small bump in $n_e l$ at $z\sim 20$ $mm$ corresponds to a higher density 
feature at the conical shield opening. 
This feature 
is the main source of self emitted radiation, while the jet itself emits less than in the case without conical 
shield (Fig.~\ref{fig:cone_emission}). The final jet is indeed free from potentially disturbing low density 
material around it, but the lower density and reduced length make it less useful for interaction experiments.

\section{Conclusions}
\label{sec:conclusions}

In this paper we presented the results of our numerical investigation on the jets produced by conical wire 
arrays on the pulsed power generator MAGPIE in London. 

First, we found that the resolution employed for the 
simulations 
plays a 
crucial 
role to correctly reproduce the experiments. Indeed, only a resolution of $\sim 100$ $\mu m$ allowed us to 
reproduce 
successfully 
the jet unstable character observed in the experiments. Coarser resolutions 
proved to be insufficient to resolve the details in the jet. However, they could be used as an approximation 
of the general trend of the physical variables in the jet.

Second, we investigated the effects of a different number of wires in the array. Higher wire numbers mean 
smoother mass distribution and smaller inter-wire gaps, therefore smaller perturbations and 
a much smoother distribution of the magnetic field. In the case of large inter-wire 
gaps however, the local magnetic field generated around each wire gives an important contribution to the 
global field. The resulting magnetic field is therefore much stronger in between the wires and extends far 
deeper into the array and nearer to the jet. 
This difference in mass distribution and magnetic field structure 
is the reason for the enhanced unstable character of the jet produced by the $16$ wires array, as compared to 
the jet in the $32$ wires array.

Lastly, we simulated the behaviour of a conical array in the presence of a conical shield. This was done to 
prevent the presence of low density plasma and flow above the array. Such plasma could potentially perturb the 
system in jet-target interaction experiments. In comparison to the arrays without conical shield, we found that 
the 
resulting jet has similar velocity but is shorter and much less dense. The overall momentum is strongly reduced, 
therefore this setup is not very suitable for jet-target interaction experiments.

This work is the basis for a deeper numerical investigation which will take into consideration, among others, 
the effects of the inclination angle on the resulting jet. Such study will be supported by, and integrated with, 
new 
sets of experiments both on  MAGPIE in London and on the ''Z'' generator in Sandia, USA. We also plan to perform 
experiments of the interaction of the jet with a gas target.

Indeed, our main focus and final goal is to connect experiments and simulations with the astrophysical scenario. 
YSO jets form deeply into molecular clouds and pierce through them during their propagation. Several details of the 
interaction between the jet and the cloud are still not completely understood. The most evident signature of 
the interaction is the strong bow shock at the head of the jet, the structure of which is the subject of numerous 
studies. Along its path, the jet entrains material from the surrounding ambient medium. The morphology and dynamic 
of these regions are of great interest for the study of the chemical evolution and turbulent behaviour of molecular 
clouds. Finally, plasma instabilities can affect the jet and potentially disrupt the flow.
Laboratory jets give us the 
opportunity to study such aspects in detail taking advantage of the accurate diagnostics available to plasma 
physicists, therefore they are a very useful tool to understand astrophysical jets.

\acknowledgments
This work was supported by the EPSRC Grant No.
EP/G001324/1 and by the NNSA under DOE Cooperative
Agreements No. DE-F03-02NA00057 and No.
DE-SC-0001063.
Part of the simulations presented here were run on the Jade supercomputer (GENCI-CINES, Paris-Montpellier, 
France), with the support of the HPC-Europa2 project funded by the European Commission - DG Research in 
the Seventh Framework Programme under grant agreement No. 228398.


\makeatletter
\let\clear@thebibliography@page=\relax
\makeatother

\end{document}